\documentclass[12pt]{article}    %%% 12pt for preprint version
\usepackage{amsfonts,amssymb,xy} %%% AMS and XyPic fonts

\makeatletter

\newcounter{sectionc}\newcounter{subsectionc}\newcounter{subsubsectionc}
\renewcommand{\section}[1] {\vspace{12pt}\addtocounter{sectionc}{1}
\setcounter{subsectionc}{0}\setcounter{subsubsectionc}{0}\noindent
	{\bf\thesectionc. #1}\par\vspace{5pt}}
\renewcommand{\subsection}[1]{\vspace{12pt}\addtocounter{subsectionc}{1}
	\setcounter{subsubsectionc}{0}\noindent
	{\bf\thesectionc.\thesubsectionc.
	 {\kern1pt \bfseries\itshape #1}}\par \vspace{5pt}}
\renewcommand{\subsubsection}[1] {\vspace{12pt}
    \addtocounter{subsubsectionc}{1}
	\noindent{\rm\thesectionc.\thesubsectionc.\thesubsubsectionc.
	 {\kern1pt \it #1}}\par\vspace{5pt}}
\newcommand{\nonumsection}[1] {\vspace{12pt}\noindent{\bf #1}
	\par\vspace{5pt}}

\newcommand{\textlineskip}{\baselineskip=15pt}  %%% for 12pt size
\newcommand{\smalllineskip}{\baselineskip=12pt} %%% for 12pt size

\def\abstracts#1#2#3{{
	\centering{\begin{minipage}{5.5in}\baselineskip=12pt\footnotesize
	\parindent=0pt #1\par
	\parindent=15pt #2\par
	\parindent=15pt #3
	\end{minipage}}\par}}

\newcommand{\bibit}{\it}
\newcommand{\bibbf}{\bf}
\renewenvironment{thebibliography}[1]
	{\frenchspacing
     \small \baselineskip=14pt
	 \begin{list}{\arabic{enumi}.}
	{\usecounter{enumi}\setlength{\parsep}{0pt}
	  \setlength{\leftmargin 20pt}{\rightmargin 0pt}   %FOR 10--99 ITEMS
	 \setlength{\itemsep}{0pt} \settowidth
	{\labelwidth}{#1.}\sloppy}}{\end{list}}

\def\@citex[#1]#2{\if@filesw\immediate\write\@auxout
	{\string\citation{#2}}\fi
\def\@citea{}\@cite{\@for\@citeb:=#2\do
	{\@citea\def\@citea{,}\@ifundefined
	{b@\@citeb}{{\bf ?}\@warning
	{Citation `\@citeb' on page \thepage \space undefined}}
	{\csname b@\@citeb\endcsname}}}{#1}}

\newif\if@cghi
\def\cite{\@cghitrue\@ifnextchar [{\@tempswatrue
	\@citex}{\@tempswafalse\@citex[]}}
\def\citelow{\@cghifalse\@ifnextchar [{\@tempswatrue
	\@citex}{\@tempswafalse\@citex[]}}
\def\@cite#1#2{{\if@cghi$\null^{#1}$\else{#1}\fi
                \if@tempswa\typeout{IJMPA warning:
	             optional citation argument ignored: `#2'} \fi}}

\def\fpage#1{\begingroup
\voffset=.3in
\thispagestyle{empty}\begin{table}[b]\centerline{\footnotesize #1}
	\end{table}\endgroup}

\def\runninghead#1#2{\pagestyle{myheadings}
\markboth{{\protect\footnotesize\it{\quad #1}}\hfill}
{\hfill{\protect\footnotesize\it{#2\quad}}}}

\def\abstracts#1#2#3{{
	\centering{\begin{minipage}{5.35truein}\baselineskip=10pt\footnotesize
	\parindent=0pt #1\par
	\parindent=15pt #2\par
	\parindent=15pt #3
	\end{minipage}}\par}}

\long\def\@makefntext#1{
\protect\noindent \hbox to 3.2pt {\hskip-.9pt
$^{{\rm\@thefnmark}}$\hfil}#1\hfill}		%CAN BE USED

\def\@makefnmark{\hbox to 0pt{$^{\@thefnmark}$\hss}}	%ORIGINAL
	
\def\ps@myheadings{\let\@mkboth\@gobbletwo
\def\@oddhead{\hbox{}
\rightmark\hfil{\footnotesize\rm\thepage}}
\def\@oddfoot{}\def\@evenhead{{\footnotesize\rm\thepage}\hfil
\leftmark\hbox{}}\def\@evenfoot{}
\def\sectionmark##1{}\def\subsectionmark##1{}}

\makeatother

\newbox\ncintdbox \newbox\ncinttbox %% noncommutative integral symbols
\setbox0=\hbox{$-$}
\setbox2=\hbox{$\displaystyle\int$}
\setbox\ncintdbox=\hbox{\rlap{\hbox
    to \wd2{\hskip-.125em\box2\relax\hfil}}\box0\kern.1em}
\setbox0=\hbox{$\vcenter{\hrule width 4pt}$}
\setbox2=\hbox{$\textstyle\int$}
\setbox\ncinttbox=\hbox{\rlap{\hbox
    to \wd2{\hskip-.125em\box2\relax\hfil}}\box0\kern.1em}

\newcommand{\ncint}{\mathop{\mathchoice{\copy\ncintdbox}%
                      {\copy\ncinttbox}{\copy\ncinttbox}%
					  {\copy\ncinttbox}}\nolimits}

\newcommand{\opname}[1]{\mathop{\mathrm{#1}}\nolimits}

\renewcommand{\a}{\alpha}           %% abbreviation for  \alpha
\renewcommand{\b}{\beta}            %% abbreviation for  \beta
\newcommand{\Dl}{\Delta}            %% abbreviation for  \Delta
\newcommand{\dl}{\delta}            %% abbreviation for  \delta
\newcommand{\eps}{\varepsilon}      %% abbreviation for  \varepsilon
\newcommand{\Ga}{\Gamma}            %% abbreviation for  \Gamma
\newcommand{\ga}{\gamma}            %% abbreviation for  \gamma
\newcommand{\la}{\lambda}           %% abbreviation for  \lambda
\newcommand{\Om}{\Omega}            %% abbreviation for  \Omega
\newcommand{\om}{\omega}            %% abbreviation for  \omega
\newcommand{\sg}{\sigma}            %% abbreviation for  \sigma
\renewcommand{\th}{\theta}          %% abbreviation for  \theta
\newcommand{\vf}{\varphi}           %% abbreviation for  \varphi

\newcommand{\A}{\mathcal{A}}        %% an algebra
\newcommand{\B}{\mathcal{B}}        %% another algebra
\newcommand{\E}{\mathcal{E}}        %% a projective module
\newcommand{\G}{\mathcal{G}}        %% spin geometry
\renewcommand{\H}{\mathcal{H}}      %% Hilbert space
\renewcommand{\L}{\mathcal{L}}      %% Lie derivative, operator ideal
\newcommand{\C}{\mathbb{C}}         %% complex numbers
\newcommand{\HH}{\mathbb{H}}        %% quaternions
\newcommand{\R}{\mathbb{R}}         %% real numbers
\newcommand{\Sf}{\mathbb{S}}        %% sphere
\newcommand{\T}{\mathbb{T}}         %% circle as a group
\newcommand{\Z}{\mathbb{Z}}         %% integers

\newcommand{\cc}{\mathbf{c}}        %% a Hochschild chain

\newcommand{\Diff}{\opname{Diff}}   %% diffeomorphism group
\newcommand{\Dom}{\opname{Dom}}     %% domain of an operator
\newcommand{\End}{\opname{End}}     %% endomorphism ring
\newcommand{\id}{\opname{id}}       %% identity map
\newcommand{\Tr}{\opname{Tr}}       %% trace of operator
\newcommand{\Onda}[1]{\widetilde{#1}} %% abbreviation for \widetilde

\newcommand{\0}{{\vphantom{\dagger}}}  %% invisible dagger

\newcommand{\del}{\partial}         %% abbreviation for  \partial
\newcommand{\op}{\oplus}            %% direct sum
\newcommand{\opp}{\circ}            %% opposite-algebra symbol
\newcommand{\ox}{\otimes}           %% tensor product
\newcommand{\semi}{\rtimes}         %% crossed product
\newcommand{\ul}{\underline}        %% for sheaves
\newcommand{\w}{\wedge}             %% exterior product
\newcommand{\x}{\times}             %% cartesian product or cross
\newcommand{\5}{\diamond}           %% for roots of trees
\newcommand{\7}{\dagger}            %% abbreviation for + symbol
\newcommand{\8}{\bullet}            %% anonymous degree
\renewcommand{\.}{\cdot}            %% anonymous variable
\renewcommand{\:}{\colon}           %% colon in  f: A -> B

\newcommand{\Coo}{C^\infty}         %% smooth functions
\newcommand{\Hoo}{\H^\infty}        %% smooth vectors
\newcommand{\ota}{\otimes_\A}       %% tensor product over \A

\newcommand{\opyop}{\oplus\cdots\oplus}    %% repeated direct sum
\newcommand{\oxyox}{\otimes\cdots\otimes}  %% repeated tensor product
\newcommand{\tsum}{\mathop{\textstyle\sum}\nolimits} %% small sum-opr

\def\<#1,#2>{\langle#1\mathbin|#2\rangle} %% Dirac notation
\def\(#1,#2){(#1\mathbin|#2)}       %% hermitian inner product

\newcommand{\dd}[1]{\frac{\partial}{\partial#1}} %% partial derivation
\newcommand{\pd}[2]{\frac{\partial#1}{\partial#2}} %% partial deriv
\newcommand{\row}[3]{{#1}_{#2},\dots,{#1}_{#3}}   %% list:  a_1,...,a_n
\newcommand{\set}[1]{\{\,#1\,\}}     %% set notation

\newcommand{\tfrac}[2]{{\textstyle\frac{#1}{#2}}}
\newcommand{\thalf}{\tfrac{1}{2}}     %% small fraction  1/2
\newcommand{\tquarter}{\tfrac{1}{4}}  %% small fraction  1/4

\newcommand{\as}{\quad\mbox{as}\enspace} %% `as' with spacing
\newcommand{\sepword}[1]{\quad\mbox{#1}\quad} %% well-spaced words

\newcommand{\twobytwo}[4]{\pmatrix{#1 & #2 \cr #3 & #4\cr}}
\newcommand{\twobytwoeven}[2]{\twobytwo{#1}{0}{0}{#2}}
\newcommand{\twobytwoodd}[2]{\twobytwo{0}{#1}{#2}{0}}

\runninghead{J. C. V\'arilly}
            {Noncommutative Geometry and Quantization}

\begin{document}

\normalsize\textlineskip
\thispagestyle{empty}
\setcounter{page}{1}

%%%\copyrightheading{}			%{Vol. 0, No. 0 (1993) 000--000}

%%%\vspace*{0.88truein}

\fpage{1}
\centerline{\bf NONCOMMUTATIVE GEOMETRY AND QUANTIZATION}
\vspace*{0.37truein}
\centerline{\footnotesize JOSEPH C. V\'ARILLY}
\vspace*{0.015truein}
\centerline{\footnotesize\it Department of Mathematics,
                             Universidad de Costa Rica}
\baselineskip=12pt
\centerline{\footnotesize\it San Pedro 2060, Costa Rica}
\vspace*{0.225truein}
%%%\publisher{(received date)}{(revised date)}
%%%\centerline{\footnotesize Received 4 December 1999}

\vspace*{0.21truein}
\abstracts{We examine some recent developments in noncommutative
geometry, including spin geometries on noncommutative tori and their
quantization by the Shale--Stinespring procedure, as well as the
emergence of Hopf algebras as a tool linking index theory and
renormalization calculations.}{}{}

\vspace*{1pt}\textlineskip

\nonumsection{Introduction}
\vspace*{-0.5pt}

\noindent
The purpose of these lectures is to survey several aspects of
noncommutative geometry, with emphasis on its applicability to
particle physics and quantum field theory. By now, it is a commonplace
statement that spacetime at short length ---or high energy--- scales
is not the Cartesian continuum of macroscopic experience, and that
older methods of working with functions on manifolds must be rethought
in order to handle granular or bubbly spacetime and the matter fields
which they support. The question as to what should replace the
continuum is an ongoing one. Noncommutative geometry (NCG) offers a
general approach to the job of describing the geometrical aspects of
nonclassical spaces. The hope is that it can provide a solid framework
for studying fundamental interactions and~QFT.

Here we consider four aspects of this general problem. In Section~1,
we discuss the noncommutative tori that have recently emerged as an
important model is string
theory.\cite{ConnesDS}$^-$\cite{SeibergWGeom} Section~2 reviews the
general structure of noncommutative geometries based on the spectral
triples of Connes.\cite{ConnesGravBreak} The third section broaches
the formulation of quantum field theories over noncommutative spaces,
using the example of a
$3$-torus.\cite{Atlas,KrajewskiW} In Section~4, we sketch how Hopf
algebras provide a link between NCG and renormalization
calculations,\cite{Kreimer}$^-$\cite{ConnesKrRH} that illuminates the
claim that ``QFT is the geometry of the world''.

\section{Noncommutative tori}

The spaces and tools of NCG can be approached from two different
points of view: either as theoretical constructs that form part of a
principled explanatory scheme, or as objects that are fortuitously
``found in Nature'' and therefore point to an (as yet incomplete)
underlying theory. A striking example of the second viewpoint is the
emergence of noncommutative tori from compactification of Matrix
models.\cite{ConnesDS}

In those models, one considers an action functional of the general
form
$$
I = \sum_{\a,\b} \Tr F_{\a\b}^2
    + 2 \sum_{i,j,\a} \Tr \Psi^i \ga_{ij}^\a \,[\nabla_\a, \Psi^j],
$$
where $\nabla$ is a connection and $F$ is its curvature. For instance,
we may ask that $\nabla_\a = X_\a$ be matrix components of some Lie
algebra representation, $F_{\a\b} = [X_\a,X_\b]$, the $\ga^\a$ being
gamma-matrices, and the $\Psi^i$ are odd variables. In the example
considered in Ref.~\citelow{ConnesDS}, the matrices $(X_\a,\Psi^i)$
are labelled by coordinates of the superspace~$\C^{10|16}$.

To compactify, we ask that at least some of the $X_\a$ variables
change only by a gauge transformation under certain fixed
translations $X_\a \mapsto X_\a + r_\a$. For two such directions, this
gives
\begin{eqnarray*}
X_0 + r_0 & = & u_0 X_0 u_0^{-1},  \\
X_1 + r_1 & = & u_1 X_1 u_1^{-1},  \\
X_\a & = & u_\b X_\a u_\b^{-1} \qquad\mbox{in all other cases}, \\
\Psi^i & = & u_\a \Psi^i u_\a^{-1} \qquad\mbox{ in all cases},
\end{eqnarray*}
for certain unitary operators $u_\a$ (to be determined). By taking
traces of the first two equations, it becomes clear that these
equations have no solutions in $N \x N$ matrices. However, there are
formal operator solutions, such as
$$
X_0 = ir_0 \dd{\phi_0} + A_0(\phi_0,\phi_1), \qquad
X_1 = ir_1 \dd{\phi_1} + A_1(\phi_0,\phi_1),
$$
and $X_\a = A_\a(\phi_0,\phi_1)$ for other~$\a$, where $\phi_0,\phi_1$
are angular variables and $u_0,u_1$ are rotations of these angles.

To get a clearer picture, notice that \textit{all} $X_\a$ and $\Psi^i$
commute with the unitary $u_1 u_0 u_1^{-1} u_0^{-1}$. If we are
looking for an irreducible solution of the equations, we can suppose
that this unitary is a scalar. Thus there is some number 
$\la = e^{2\pi i\th}$ of absolute value $1$, such that
\begin{equation}
u_1 u_0 = e^{2\pi i\th} u_0 u_1.
\label{eq:NCT2-comm-rel}
\end{equation}
Now, the point is that nothing we have said so far demands that $\la$
be equal to~$1$ (that is, $\th = 0$), so we can perfectly well suppose
that it is not. Indeed, if $\th = 0$, then $u_0$ and $u_1$ can be
taken as coordinate rotations on the ordinary torus~$\T^2$. For any
other (non-integer) value of~$\th$, $u_0$ and $u_1$ become
``coordinates'' for a certain noncommutative space, namely, the NC
torus~$\T_\th^2$.

When compactifying in more than two dimensions, we obtain more
relations, of the form
\begin{equation}
u_k u_j = e^{2\pi i\th_{jk}} u_j u_k,
\label{eq:NCTn-comm-rel}
\end{equation}
where $\th = [\th_{jk}]$ is now a real skewsymmetric matrix.

\newpage

\subsection{The NC torus as a noncommutative space}

The last statement is to be interpreted as follows. Noncommutative
topology replaces a locally compact space of points $Y$ by its algebra
$C_0(Y)$ of continuous functions vanishing at infinity; this is a
commutative $C^*$-algebra. No information is lost (or gained) in
passing from the space to the algebra, and the process is reversible:
this is the content of the Gelfand--Na\u{\i}mark
theorem.\cite{GelfandN} For instance, the space is compact if and only
the algebra has a unit element, the space is disconnected if and only
if the algebra contains nontrivial idempotents, and so on. To deal
with gauge potentials and suchlike, it is often better to work instead
with the dense subalgebra of smooth functions. Finally, we abandon
points by discarding the commutativity property and by calling any
$C^*$-algebra a ``noncommutative space''; or, if smoothness remains
important, we work with certain dense subalgebras called
pre-$C^*$-algebras.

From this point of view, the NC torus $\T_\th^2$ is just a certain
dense subalgebra of the $C^*$-algebra generated by two unitaries
subject only to the relation~(\ref{eq:NCT2-comm-rel}). Notice that
this is not the notorious ``quantum plane'', since the requirement of
unitarity is met by four more relations: 
$u_0^\0 u_0^\7 = u_0^\7 u_0^\0 = 1$,
$u_1^\0 u_1^\7 = u_1^\7 u_1^\0 = 1$, which are not asked of the
quantum plane.

The $n$-dimensional noncommutative tori are to be regarded as
quantizations of the ordinary torus, whose algebra is
$\Coo(\T^n) =: \T_0^n$. Indeed, noncommutativity is achieved by
replacing the ordinary product of functions on~$\T^n$ with the Moyal
product: see, for instance, Ref.~\citelow{WeinsteinTheta} or our
Ref.~\citelow{Polaris}. A quick-and-dirty way to do this is by
describing elements of $\T_\th^n$ as ``noncommutative Fourier
series''.

For definiteness, we take $n = 3$ and let $\th$ be a real
skewsymmetric $3 \x 3$ matrix. The $3$-torus is generated by three
unitaries $u_1,u_2,u_3$, subject to the commutation relations
(\ref{eq:NCTn-comm-rel}) and no others. Introduce the unitary Weyl
elements\cite{Atlas,Bedos}
$$
u^r := \exp\{\pi i(r_1\th_{12}r_2 + r_1\th_{13}r_3 + r_2\th_{23}r_3)\}
           \, u_1^{r_1} u_2^{r_2} u_3^{r_3},
$$
for each $r \in \Z^3$; the coefficient is chosen so that
$(u^r)^* = u^{-r}$ in all cases. They obey the product rule
\begin{equation}
u^r\,u^s = \la(r,s)\, u^{r+s},  \qquad
 \la(r,s) := \exp\bigl\{-\pi i\tsum_{j,k} r_j \th_{jk} s_k\bigr\}.
\label{eq:NCT-prod}
\end{equation}
Notice that $|\la(r,s)| = 1$ and $\la(r,\pm r) = 1$ by skewsymmetry
of~$\th$. This $\sg$ is in fact a $2$-cocycle for the abelian group
$\Z^3$ and the $C^*$-algebra $C^*(\Z^3,\sg)$ is generated by the $u^r$
subject to this product rule is called a twisted group $C^*$-algebra.
The NC torus $\T_\th^3$ is the dense subalgebra consisting of all
Fourier series
$$
\T_\th^3 := \set{a = \tsum_r a_r\, u^r : a_r \to 0 \mbox{ rapidly}},
$$
where rapid decrease of the coefficients means that
$(1 + |r|^2)^k \,|a_r|^2$ is bounded for all $k = 1,2,3,\dots$. In the
commutative case $\th = 0$, this condition gives
$\T_0^3 \simeq \Coo(\T^3)$.

This algebra is naturally represented on a certain Hilbert space,
using an old trick (the GNS construction) that requires a faithful
state on the algebra. In fact, there is one such state which is also a
\textit{trace}:
$$
\tau\bigl( \tsum_r a_r u^r \bigr) := a_0.
$$
By completing $\T_\th^3$ in the Hilbert norm
$$
\|a\|_2 := \sqrt{\tau(a^*a)} = \bigl( \tsum_r |a_r|^2 \bigr)^{1/2},
$$
we get a Hilbert space $\H_\tau$. We write $\ul{c}$ for an element
$c \in \T_\th^3$ regarded as a vector in~$\H_\tau$. Then the GNS
representation of $\T_\th^3$ is just
$$
\pi(a) : \ul{c} \mapsto \ul{ac}.
$$
In other words, $\T_\th^3$ acts on $\H_\tau$ by left multiplication
operators.

The Lie group $\T^3$ acts by rotations on the algebra $\T_\th^3$, as
follows: for each~$r$,
$(z_1,z_2,z_3)\.u^r := z_1^{r_1} z_2^{r_2} z_3^{r_3} \,u^r$. The
trace~$\tau$ is invariant under this action; indeed, $\tau$ picks out
the only rotation-invariant component of an element of~$\T_\th^3$.

There is a special \textit{antiunitary} operator $J_0$ on~$\H_\tau$,
given by
$$
J_0(\ul{a}) := \ul{a^*}.
$$
This is the Tomita conjugation\cite{KadisonR} determined by the cyclic
and separating vector~$\ul{1}$ for the representation~$\pi$;
clearly, $J_0^2 = 1$. The operator
$$
\pi^\opp(b) := J_0 \pi(b^*) J_0
 : \ul{c} \mapsto J_0 \ul{b^*c^*} = \ul{cb}
$$
is a \textit{right} multiplication by the element~$b$, and is an
antirepresentation of $\T_\th^3$. Equivalently, $\pi^\opp$ is a true
representation of the ``opposite algebra'' of~$\T_\th^3$, obtained by
reversing the product. A glance at~(\ref{eq:NCTn-comm-rel}) shows
that this opposite algebra is $\T_{-\th}^3$. Life is simpler if we
forget to write the $\pi$'s; the commutativity of left and right
multiplications is then expressed as
\begin{equation}
[a, J_0 b^* J_0] = 0  \sepword{for all} a,b.
\label{eq:Jzero-LRcomm}
\end{equation}

Differential calculus on tori begins with the partial derivatives
$$
\dl_j\bigl(\tsum_r a_r u^r\bigr) := 2\pi i \tsum_r r_j\, a_r u^r
  \qquad  (j = 1,2,3).
$$
To see why they are partial derivatives, pretend that $\th = 0$, so
that the $u^r = \exp\{2\pi i(r_1\phi_1 + r_2\phi_2 + r_3\phi_3)\}$ are
a basis for an (ordinary) Fourier expansion on~$\T^3$. As operators on
the algebra, the $\dl_j$ are symmetric derivations:
$$
\dl_j(ab) = (\dl_j a)b + a(\dl_j b),  \qquad \dl_j(a^*) = (\dl_j a)^*,
$$
and they satisfy $\tau(\dl_j a) = 0$.

With the canonical trace $\tau$ and the partial derivatives, we build
up certain rotation-invariant multilinear forms on~$\T_\th^3$. Besides
$\tau$ itself, we find
\begin{eqnarray*}
\psi_j(a,b) & := & \tau(a\,\dl_j b),  \\
\vf_{jk}(a,b,c) & := & \frac{1}{2\pi i}
     \, \tau(a\,\dl_j b\,\dl_k c - a\,\dl_k b\,\dl_j c)  \\
\om(a,b,c,d) & := & - \frac{1}{4\pi^2}\, \eps^{ijk}
     \, \tau(a\,\dl_i b\,\dl_j c \,\dl_k d).
\end{eqnarray*}
When $\th = 0$, these forms correspond to the homological structure of
the ordinary $3$-torus: think of the vertex, the edges, the faces and
the interior of a periodic box in $3$-space. For instance,
$\psi_j(a,b)$ is matched with the line integral of the $1$-form
$a\,db$ over the $j$th edge of the box. A deep theorem of
Connes\cite{ConnesNCDiffG} relates the de~Rham homology of a
(commutative) manifold $M$ to the cyclic cohomology of the algebra
$\Coo(M)$: after factoring out certain redundancies, the resulting
``periodic cyclic cohomology'' of $\Coo(M)$ is isomorphic to the
de~Rham homology of~$M$. It turns out that $\tau$, $\psi_j$,
$\vf_{jk}$ and $\om$ are cyclic cocycles, of respective degrees $0$,
$1$, $2$ and~$3$, for the torus $\T_\th^3$, independently of~$\th$.

There is also an algebraic counterpart of de~Rham cohomology. There is
a theory of cyclic homology of algebras,\cite{BrodzkiandLoday} but we
only need the (less complicated) Hochschild homology, and only in the
top degree. The chains for this homology theory are multiple tensor
products of algebra elements. Here, for example, is the algebraic
counterpart of the \textit{volume form} on a $3$-torus:
$$
\cc := \frac{1}{6(2\pi i)^3}\, \eps^{ijk}
         u_i^{-1}u_j^{-1}u_k^{-1} \ox u_k \ox u_j \ox u_i.
$$
The boundary operator $b$ collapses one tensor product to a
multiplication, yielding the following alternating sum:
$$
b(t \ox u \ox v \ox w)
 = tu \ox v \ox w - t \ox uv \ox w + t \ox u \ox vw - wt \ox u \ox v,
$$
and an easy calculation\cite{Atlas} shows that $b\,\cc = 0$, so
that $\cc$ is a Hochschild $3$-cycle.

\subsection{Spin geometry on the NC torus}

Geometry now enters the picture through the \textit{Dirac operator}
$$
D = -i\,\ga^\mu \dl_\mu
$$
where the (Euclidean) gamma matrices satisfy
$\ga^\mu \ga^\nu + \ga^\nu \ga^\mu = 2\,\dl^{\mu\nu}$. When $n = 2m$
or $2m + 1$, these act on a vector space of dimension~$2^m$. The
\textit{spinor space} is then
$$
\H := \H_\tau \op \H_\tau \opyop \H_\tau  \quad\mbox{($2^m$ times)}.
$$
In the cases $n = 2$ or~$3$, there are two copies of $\H_\tau$ and
the gamma matrices are just the standard Pauli matrices. For $n = 3$,
\begin{equation}
D := -i(\sg_1\,\dl_1 + \sg_2\,\dl_2 + \sg_3\,\dl_3)
 = -i\pmatrix{\dl_3 & \dl_1 - i\dl_2 \cr \dl_1 + i\dl_2 & -\dl_3 \cr}.
\label{eq:Dirac-NCT}
\end{equation}
For the case $n = 2$, there is no $\dl_3$ and the diagonal entries
are replaced by zeros, so $D$ is an odd matrix. Indeed, the
grading operator $\ga^3 := -i\,\ga^1\ga^2$ anticommutes with $D$
when $n = 2$. For $n = 3$, there is no grading available. When
$\th = 0$, we recover the well known Dirac operators on tori (with the
standard flat metric and untwisted spin structure).

The Riemannian distance on the ordinary torus $\T^n$ is determined by
the Dirac operator, through the formula\cite{ConnesMetric}
$$
d(p,q) = \sup\set{ |a(p) - a(q)|
                   : a \in C(\T^n),\ \|[D,a]\| \leq 1}.
$$
See Ref.~\citelow{Polaris} or Section~VI.1 of Ref.~\citelow{Book}
for a proof of that; the formula works because $\|[D,a]\|$ is
the sup norm of  the gradient of~$a$. The formula also makes sense, as
a definition,  for noncommutative tori, with $C(T^n)$ replaced by the
$C^*$-algebra  completion of $\T_\th^n$ and $p,q$ interpreted as pure
states of  this $C^*$-algebra. Unfortunately, that is of little use
since the  state space is quite complicated. The issues of using the
distance  formula for noncommutative algebras are thoroughly discussed
by  Rieffel in Ref.~\citelow{RieffelMetric}.

The \textit{charge conjugation} operator on~$\H$ is given by
$$
C := -i\ga^2 \ox J_0 = \pmatrix{0 & -J_0 \cr J_0 & 0 \cr},
$$
for $n = 2$ or~$3$. For higher $n$, the first factor in~$C$ is a
suitable product of gamma matrices that makes $C$ antiunitary and
satisfies $C \ga^\mu C^{-1} = -\ga^\mu$. In general, $C^2 = \pm 1$,
with a dimension-dependent sign. In the commutative case, the factor
$J_0$ reduces to complex conjugation of functions.

By combining several of these ingredients, we can represent Hochschild
chains on the spinor spaces. An $r$-chain with values in
$\T_\th^n \ox \T_{-\th}^n$ is a sum of terms of the form
$(a \ox b^\opp) \ox a_1 \oxyox a_r$ where $b^\opp = \pi^\opp(b)$ lies
in the opposite algebra; such a term is represented by the operator
$$
\pi_D((a \ox b^\opp) \ox a_1 \oxyox a_n)
 := a C b^* C^{-1} \,[D,a_1] \dots [D,a_n].
$$
When $b = 1$, we omit it. For example, if $n = 3$, we see that
$[D,u_j] = 2\pi\,\sg_j$, so the volume form is represented by
$$
\pi_D(\cc) = \frac{(2\pi)^3}{6(2\pi i)^3}
  \eps^{ijk}\,\sg_k \sg_j \sg_i
 = \frac{(2\pi)^3}{6(2\pi i)^3}\, (-6i) = 1.
$$
When $n = 4$, the analogous calculation gives
$\pi_D(\cc) = \ga_5$, from the product of four gamma
matrices. In general, the volume form is represented by~$1$ or by the
grading operator, according as $n$ is odd or even.

The dimension may also be extracted from the Dirac operator, by
examining the growth of its spectrum. Ignoring zero modes (of which
there are only a finite number), we find that $D^{-1}$ is a compact
selfadjoint operator, so that both $D$ and $|D| := \sqrt{D^\7 D}$ have
discrete spectra. If $s_k(D)$ denotes the $k$th singular value of~$D$
(i.e., the $k$th eigenvalue of~$|D|$) in increasing order, there is
one and only one integer~$n$ for which
$$
\sum_{k\leq N} s_k(D)^{-n} \sim C \log N  \as  N \to \infty.
$$
This $n$ is the \textit{classical dimension} of the NC torus. Write
$C =: \ncint |D|^{-n}$ for the coefficient of logarithmic divergence.
What happens is that, just as in calculations of the dimension of
fractals, there is one critical value of~$n$ with
$0 < \ncint |D|^{-n} < \infty$, while $\ncint |D|^{-s}$ is zero for
$s > n$ and is infinite for $s < n$.

We have enough information now to compute the classical dimension
of~$\T^3$ directly. First of all,
$D^2 = - (\sg\.\dl)^2 = (-\dl_1^2 - \dl_2^2 - \dl_3^2)$ can be
diagonalized in the orthonormal basis of $\H$ given by
$$
\psi_r^+ := \pmatrix{\ul{u^r} \cr 0 \cr},  \quad
\psi_r^- := \pmatrix{0 \cr \ul{u^r} \cr},  \quad  r \in \Z^3.
$$
Indeed, $D^2 \psi_r^\pm = 4\pi^2|r|^2 \psi_r^\pm$, and therefore
$|D| \psi_r^\pm = 2\pi|r| \psi_r^\pm$. There are two zero modes,
$\psi_0^+$ and $\psi_0^-$, which we ignore when dealing with
$|D|^{-1}$. We compute
\begin{equation}
\ncint |D|^{-s} = 2 \lim_{R\to\infty} \frac{1}{3\log R}
                \sum_{1\leq |r|\leq R} (2\pi|r|)^{-s}
 = \lim_{R\to\infty} \frac{2}{3\log R}
     \int_1^R \frac{4\pi\rho^2\,d\rho}{(2\pi\rho)^s},
\label{eq:cldim-NCTthree}
\end{equation}
which is zero for $s > 3$, diverges for $s < 3$ and equals $1/3\pi^2$
for $s = 3$; so indeed the classical dimension is~$3$, as expected.

The geometrical apparatus outlined here (Dirac operator, spinor space,
cyclic cocycles, volume form, charge conjugation, dimension) of course
works the same in the commutative case; to be precise, this is the
geometry of compact spin manifolds. In Refs.~\citelow{Polaris}
and \citelow{Portia}, the example of the Riemann sphere is done
in full detail, and the general theory is laid out in Part~III of
Ref.~\citelow{Polaris}. In other words, the classical geometry of
Riemannian spin manifolds can be rewritten in a purely operatorial
language; but in that new language, many other geometries also appear,
that may be called spin geometries over noncommutative spaces.

\section{Rules and procedures for NC geometries}

We now explain more systematically how the various pieces of the
geometrical apparatus fit together. We also need to see how it
can serve as the basis for other models of physical interest.
One of the most striking of these was a phenomenological
Yang--Mills--Higgs model put forward by Connes and
Lott,\cite{Book,ConnesL,ChamseddineC} in order to
incorporate symmetry breaking at the geometrical level,\cite{DuBoisVKM}
and later developed by several
others.\cite{ScheckEtAl}$^-$\cite{Chamseddine} A detailed
review of this approach to the Standard Model is given in our
Ref.~\citelow{Cordelia}.

\newpage

\subsection{The ground rules for spin geometries}

We define a noncommutative spin geometry by a list of \textit{terms}
and \textit{conditions}.

The terms form a package $\G = (\A,\H,D,C,\chi)$, where $(\A,\H,D)$ is
a \textit{spectral triple}.\cite{ConnesReal} This means that $\A$ is
an algebra, represented as operators on a Hilbert space $\H$, and
that $D$ is an (unbounded) selfadjoint operator on~$\H$ such that
$\ker D$ is finite-dimensional and $D^{-1}$ is compact on
$\H \ominus \ker D$, and also that $[D,a]$ is bounded for each
$a \in \A$. The operator $C$ is an antiunitary conjugation on~$\H$
such that $b \mapsto C b^* C^{-1}$ is a representation of the
opposite algebra $\A^\opp$ which commutes with $\A$, that is,
$[a, C b^* C^{-1}] = 0$ for all $a,b \in \A$. Finally, for $\chi$
there are two cases: in the odd case $\chi = 1$, and in the even case
$\chi$ is a grading operator, that is, a selfadjoint operator
such that $\chi^2 = 1$. In the even case, the algebra acts evenly,
$\chi a = a \chi$, while the operator $D$ is odd, $\chi D = - D \chi$.

There are seven conditions to satisfy,\cite{ConnesGravBreak} which we
now list.

\begin{enumerate}
\item
\textit{Classical dimension}: there is a nonnegative
integer~$n$, that is odd or even according as $\chi = 1$ or not, such
that $\sum_{k\leq N} s_k(D)^{-n} \sim C \log N$ as $N \to \infty$,
with $0 < C < \infty$. We write $\ncint |D|^{-n} := C$.
If $\A$ and $\H$ are finite-dimensional, we set $n = 0$.

\item
\textit{Regularity}: the bounded operators $a$ and $[D,a]$, for any
$a \in \A$, belong to the domain of smoothness
$\bigcap_{k=1}^\infty \Dom(\dl^k)$ of the derivation
$\dl(T) := [|D|,T]$.

\item
\textit{Finiteness}: the space of smooth vectors
$\Hoo := \bigcap_{k=1}^\infty \Dom(D^k)$ is a
finite projective left module over the pre-$C^*$-algebra $\A$. It
carries an $\A$-valued inner product $\(\.,\.)$ implicitly defined by
$\ncint\, \(\phi,\psi)\,|D|^{-n} = \<\phi,\psi>$.

\item
\textit{Reality}: the conjugation $C$ satisfies $C^2 = \pm 1$,
$CD = \pm DC$, and $C\chi = \pm\chi C$ in the even case, where the
signs are given by the following tables:
$$
\mbox{\footnotesize\smalllineskip
\mbox{
\begin{tabular}{ccccc}
\hline
\rule[-4pt]{0pt}{14pt} $n \bmod 8$         & $0$ & $2$ & $4$ & $6$  \\
\hline
\rule[-4pt]{0pt}{14pt} $C^2 = \pm 1$       & $+$ & $-$ & $-$ & $+$  \\
\rule[-4pt]{0pt}{14pt} $CD = \pm DC$       & $+$ & $+$ & $+$ & $+$  \\
\rule[-4pt]{0pt}{14pt} $C\chi = \pm\chi C$ & $+$ & $-$ & $+$ & $-$  \\
\hline
\end{tabular}}\qquad\quad
\mbox{
\begin{tabular}{ccccc}
\hline
\rule[-4pt]{0pt}{14pt} $n \bmod 8$   & $1$ & $3$ & $5$ & $7$  \\
\hline
\rule[-4pt]{0pt}{14pt} $C^2 = \pm 1$ & $+$ & $-$ & $-$ & $+$  \\
\rule[-4pt]{0pt}{14pt} $CD = \pm DC$ & $-$ & $+$ & $-$ & $+$
\\ \hline
\rule[-4pt]{0pt}{14pt}
\end{tabular}}}
$$

\item
\textit{First order}: $[[D,a], Cb^*C^{-1}] = 0$ for all $a,b \in \A$.

\item
\textit{Orientation}: there is a Hochschild $n$-cycle
$\cc \in (\A \ox \A^\opp) \ox \A^{\ox n}$, $b\,\cc = 0$, whose
representative on~$\H$ satisfies $\pi_D(\cc) = \chi$.

\item
\textit{Poincar\'e duality}:
The Fredholm index of the operator $D$ yields a nondegenerate
intersection form on the $K$-theory of the algebra $\A$.

\end{enumerate}

We refer to Ref.~\citelow{ConnesGravBreak} and Section~10.5 of
Ref.~\citelow{Polaris} for a full discussion of these conditions
and their implications. Here we just make a few remarks.

To produce a ``compact NC space'', we demand that $\A$ have a
unit~$1$ (think of the constant function~$1$ on a manifold), and that
$D^{-1}$ be compact outside $\ker D$. This may be weakened in the
``locally compact'' (i.e., nonunital) case by asking instead that
$(1 + D^2)^{-1/2}$ be compact. We no longer need to suppose that $D$
is any sort of Dirac operator.

The dimension condition is obtained from a version of Weyl's formula
linking the dimension and volume of a compact manifold to the growth
of the eigenvalues of the Laplacian. For spin manifolds, the
Lichnerowicz formula $D^2 = \Dl + \tquarter s$ says that $D^2$ is a
generalized Laplacian on spinors.

The regularity condition, in the commutative case, expresses the
smoothness of the coordinate functions~$a$; this can be seen by
working out the symbols of $\dl^k([D,a])$ with pseudodifferential
calculus. Another method, developed by Rennie,\cite{Rennie} also gives
the smoothness of the functions. For the NC tori, the regularity
condition imposes the fast decrease of the Fourier series
coefficients.

The finiteness condition asks that the space $\Hoo$ of smooth vectors
be either $\A^N$, a direct sum of several copies of~$\A$ (so operators
and vectors have the same smoothness properties), or at least of the
form $p\A^N$ for some projector $p \in M_N(\A)$. For the NC tori
example with $\A = \T_\th^n$, we took $\Hoo = \A^N$ where $N = 2^m$
when $n = 2m$ or $2m + 1$. In the commutative case, $\Hoo$ is the
space of smooth sections of the spinor bundle.

The tables for the reality condition signal an underlying action of
the real Clifford algebra of~$\R^n$, complete with mod-$8$
periodicity, that indeed gives rise to the charge conjugation
operator~$C$: for the full story, see Chapter~9 of
Ref.~\citelow{Polaris}. In short, the sign tables arise from the
product rules for gamma matrices.

The Poincar\'e duality condition is a reinterpretation, in $K$-theory
language, of the usual pairing of differential forms of complementary
degrees on an oriented manifold: see Section~VI.4 of
Ref.~\citelow{Book}.

Two geometries $\G_1$ and $\G_2$ give rise to a product geometry $\G$,
with $\A = \A_1 \ox \A_2$, $\H = \H_1 \ox \H_2$, 
$\chi_1 \ox 1 + 1 \ox \chi_2$ if, say, $\G_1$ is even. We take 
$D = D_1 \ox 1 + \chi_1 \ox D_2$. The product conjugation is worked
out on a case-by-case basis.\cite{Vanhecke}

\subsection{Finite geometries}

Let us look at a ``baby geometry'' of classical dimension zero, where
$\A$ is a finite-dimensional matrix algebra (stable under taking
Hermitian conjugates), acting on a finite-dimensional Hilbert space
$\H$. Then $D$ is a matrix, too, with a finite number of eigenvalues.

This means that any scheme of approximation of a higher-dimensional
space by a set of finite matrix algebras is not a straightforward
matter. A popular model is the so-called fuzzy
sphere,\cite{MirandaEtAl} where the algebra $\Coo(\Sf^2)$ is
approximated by algebras $M_{2j+1}(\C)$ obtained from the Moyal
product on the phase space~$\Sf^2$. The issue is how to control the
limit $j \to \infty$ when the classical dimension jumps from~$0$
to~$2$; it is fair to say that this problem is still open.

It is instructive, even so, to consider the ``two-point space''
over the (commutative!) algebra $\A = \C^2$, with $\H = \C^2$ also.
We use $\chi = \twobytwoeven{1}{-1}$ as the grading operator, so
that $\A$ must act by diagonal $2 \x 2$ matrices. Since $n = 0$, the
tables give $C^2 = +1$ and $C\chi = \chi C$, so $C$ must be just
complex conjugation. The selfadjoint odd operator $D$ is of the form
\begin{equation}
D = \twobytwoodd{m^*}{m}  \sepword{for some}  m \in \C.
\label{eq:baby-Yuk}
\end{equation}
But $D = CDC = \twobytwoodd{m}{m^*}$, so $m$ is real. We may as well
suppose that $m > 0$. The distance between the two points (the pure
states of $\A$) is found from
$$
[D,a] = \twobytwoodd{-m(a_1 - a_2)}{m(a_1 - a_2)},
$$
so that $[D,a]^2 = -m^2(a_1 - a_2)^2\,1_2$, and thus
$\|[D,a]\| = m|a_1 - a_2|$. The maximum value of $|a_1 - a_2|$ when
$\|[D,a]\| \leq 1$ equals~$1/m$. This already indicates that $m$ is
an inverse distance, so it can be regarded as a mass.

A more elaborate proposal, arising from the Connes--Lott
models,\cite{ConnesL,KastlerAccounts,SchueckerEtAl,Cordelia} is to use
a baby geometry to stand for the Yukawa terms in the Standard Model
Lagrangian (at the classical level). This can be thought of ---when
combined with the spacetime geometry--- as a useful test case of what
a realistic NC geometry would look like.

The Hilbert space for the ``baby Yukawa'' model has a basis
labelled by Weyl fermions. The conjugation $C$ exchanges the
particle and antiparticle subspaces, and reduces $D$ to the act on
each of these separately. We can again split with $\chi$, according
to chirality; on the particle subspace, $D$ then looks
like~(\ref{eq:baby-Yuk}), where now $m$ is a matrix of lepton and
quark masses, incorporating the Cabibbo--Kobayashi--Maskawa mixing.

The algebra of this model is dictated by the gauge symmetry of the
Standard Model. We would like a matrix algebra whose unitary group is
$U(1) \x SU(2) \x SU(3)$, but there is none available. The best we can
manage is to use the local gauge group $U(1) \x SU(2) \x U(3)$, and
later eliminate the extra $U(1)$ factor; it turns out that this can
be done by anomaly cancellation.\cite{Chiron} The corresponding
algebra is\cite{Book}
$$
\A = \C \op \HH \op M_3(\C),
$$
where $\HH$ denotes the quaternions. (Had we tried instead to use
$\HH \op M_3(\C)$, whose unitary group is $SU(2) \x U(3)$, we would
have obtained a model whose leptons are not colour singlets.)

The final step in building the baby Yukawa model is the choice of the
representation of $\A$ on the Hilbert space; this involves a delicate
balancing, with different prescriptions for the lepton and quark
sectors. For the details of the construction, see
Ref.~\citelow{Cordelia}.

A complete analysis of the finite spectral triples that satisfy the
requirements for a noncommutative spin geometry is
available.\cite{PaschkeEtAl,KrajewskiThesis} One finds,
in essence, that when Poincar\'e duality is taken into account, the
baby Yukawa model is the simplest case that satisfies all the above
conditions.

\subsection{Gauging the spin geometries}

Gauge potentials are ``$1$-forms'' that give Hermitian operators:
$$
A := \tsum_j a_j\,[D,b_j]  \sepword{with}  A^* = A.
$$
By $1$-forms we mean finite sums of the indicated kind. Since
$a\,[D,b]\,c = a\,[D,bc] - ab\,[D,c]$, such $1$-forms make up an
$\A$-bimodule of operators on~$\H$, which we call $\Om_D^1$. The
gauging rule for the operator $D$ is then
$$
D \mapsto D + A + CAC^{-1}.
$$
Fermionic action terms are schematically like
$I(\Psi) = \<\Psi, (D + A + CAC^{-1})\Psi>$.

If we apply to $\A$ an inner automorphism $a \mapsto uau^*$, where $u$
is a unitary element of~$\A$, there is a corresponding unitary
operator on~$\H$ given by $U := u C u C^{-1}$. Since $C u C^{-1}$
commutes with $\A$, the operator $U$ implements the same inner
automorphism. Moreover, $U$ commutes with $C$ (by design) and
with~$\chi$. The upshot is that the package $\G = (\A,\H,D,C,\chi)$
is unitarily equivalent to $(\A,\H, UDU^{-1}, C,\chi)$, and so may be
regarded as the same spin geometry. We now compute
\begin{eqnarray*}
UDU^{-1} & = & CuC^{-1}u\,D\,u^* Cu^*C^{-1}
           =   CuC^{-1} (D + u\,[D,u^*]) Cu^*C^{-1}  \\
& = & CuC^{-1} D Cu^*C^{-1} + u\,[D,u^*]
  =   D + CuC^{-1}\,[D,Cu^*C^{-1}] + u\,[D,u^*]  \\
& = & D + u\,[D,u^*] + Cu\,[D,u^*]C^{-1}.
\end{eqnarray*}
What this means is that the gauge potential $A := u\,[D,u^*]$ is
trivial, in that it does not alter the geometry.

For a more general selfadjoint $A \in \Om_D^1$, the recipe
$$
\nabla c =: [D,c] + A c
$$
defines a covariant derivative on~$\A$ with values in $\Om_D^1$. We
then get a new selfadjoint operator $\Onda D$ on~$\H$ from the Leibniz
rule
\begin{eqnarray*}
\Onda D(b\psi) & := & (\nabla b)\psi + b\,D\psi
                                     + b C(\nabla 1)C^{-1}\psi  \\
& = & [D,b]\psi + A b\psi + b\,D\psi + b CAC^{-1} \psi  \\
& = & (D + A + CAC^{-1})\, b\psi.
\end{eqnarray*}
The new geometry is generally not unitarily equivalent to the
original one, so we have arrived at a wider notion of gauge
equivalence of geometries.

In fact, one can go much further, generalizing the operator
$\nabla\: \A \to \Om_D^1$ by introducing connections on finitely
generated projective (right) $\A$-modules, that play the role of
vector bundles in NCG. Given such a module $\E$, a connection is a
linear map $\nabla\: \E \to \E \ota \Om_D^1$ that satisfies a Leibniz
rule
$$
\nabla(sa) = (\nabla s)a + s \ox [D,a].
$$
Working with this connection involves changing the Hilbert space to
$\E \ota \H \ota \overline{\E}$ and the algebra to $\B := \End_\A\E$,
namely, the operators on~$\E$ that commute with the right action
of~$\A$. Such a relation among algebras is called Morita equivalence.
We refer to Ref.~\citelow{Portia} for more information on that, and
to Ref.~\citelow{LandiBook} for a full discussion of connections. The
point is that Morita equivalence of algebras is directly linked to
gauge equivalence of geometries.

\section{QFT on a noncommutative space}

We now consider, albeit briefly, what we can learn about quantum field
theory by using the technology of noncommutative spaces and
geometries. The emergence of the NC tori in compactified Matrix
theories shows that, in principle, one should be prepared to ``go
noncommutative'' at the outset. Indeed, there has been a recent
revival of interest in the idea of noncommuting spacetime coordinates.
Doplicher, Fredenhagen and Roberts\cite{DoplicherFR} provided strong
evidence for a Moyal product structure on the spacetime variables
(noncommutative $\R^4$). Recently, Seiberg and Witten, pulling various
strands together, have discussed the pervasive presence of
noncommutative tori in string theory (see Ref.~\citelow{SeibergWGeom}
and references therein).

The first to follow up the suggestion that noncommuting coordinates
might help to mollify UV divergences was Filk,\cite{Filk} who analyzed
how the Moyal product affected path-integral divergences; his
analysis, based on general topological properties of the Feynman
graphs, found no overall improvement in the UV behaviour. Our own
approach\cite{Atlas} starts with canonical quantization of fermions,
treating the gauge field as an external classical source; we also
found no better overall UV behaviour. By quantizing the nonlinear
theory by the dual way of treating the fermions with Fock space
techniques, and the gauge bosons by functional integration, this
conclusion can be shown to be generally valid. Other recent
investigations\cite{KrajewskiW,ChaichianEtAl} reinforce this view. The
question is why this ``no-go'' theorem should hold.

\subsection{Quantization of NC spaces}

We illustrate the project by setting up the free Dirac equation on a
noncommutative $3$-torus. Writing $p_j := -i\,\dl_j$, we recall
from~(\ref{eq:Dirac-NCT}) that
$$
D = -i\,\sg\.\dl = \sg\.p,
$$
so that the solutions of the equations
$$
(i\del_t - \sg\.p) \psi_R = 0,  \qquad  (i\del_t + \sg\.p) \psi_L = 0
$$
represent Weyl neutrinos on the spinor space
$\H = \H_\tau \op \H_\tau$. If one wishes, one can introduce a mass
term which couples these neutrinos:
$$
(i\del_t - \sg\.p) \psi_R = m\psi_L,  \qquad
(i\del_t + \sg\.p) \psi_L = m\psi_R,
$$
or, more compactly,
\begin{equation}
\twobytwoodd{i\del_t + \sg\.p}{i\del_t - \sg\.p}\psi = m\psi,
  \sepword{with}  \psi := \pmatrix{\psi_R \cr \psi_L \cr}.
\label{eq:Dirac-eqn}
\end{equation}
Write $E(p) := (p_1^2 + p_2^2 + p_3^2 + m^2)^{1/2}$. This is the
positive square root of a positive operator on~$\H_\tau$, and we can
introduce two more positive operators,
$$
(\sg p) \equiv \sg^\mu p_\mu := E + \sg\.p,  \qquad
(\bar\sg p)  := E - \sg\.p.
$$
Their positive square roots, in turn, are given by
$$
\sqrt{(\sg p)} \mbox{ or } \sqrt{(\bar\sg p)} := \frac{1}{\sqrt{2}}
  \,\{(E + m)^{1/2} \pm (E - m)^{1/2}\,\sg\.p\}.
$$
The equation~(\ref{eq:Dirac-eqn}) then has ``plane-wave'' solutions of
the form
$$
\psi_{E,r} = e^{-iEt} u^r \pmatrix{\sqrt{(\bar\sg r)}\,\xi \cr
                                   \sqrt{(\sg r)} \,\xi \cr},
$$
with $r \in \Z^3$ and $\xi \in \C^2$, $|\xi| = 1$.

In brief, the usual Dirac equation calculations go through in
$\R_t \x \T_\th^3$, on replacing positive functions with positive
operators.

The \textit{phase} of the (free) Dirac operator is
$F := D|D|^{-1} = D(D^2)^{-1/2}$. On each two-dimensional subspace
of~$\H$ spanned by $\{\psi_r^+,\psi_r^-\}$, for
$r = (r_1,r_2,r_3) \in \Z^3$, we can express $D$ and $F$ in terms of
Pauli matrices:
\begin{equation}
D = 2\pi\, \sg\.r,  \qquad  |D| = 2\pi\,|r|,  \qquad
F = \frac{\sg\.r}{|r|}.
\label{eq:DandF-block}
\end{equation}
The eigenvalues are then the same as for the ordinary Dirac operator
on the ordinary torus (with untwisted boundary conditions).

In order to proceed to Fock space, we split the ``one-particle space''
as $\H = \H^+ \op \H^-$, in positive- and negative-frequency solutions
of the free Dirac equation. That is to say, the grading operator for
this splitting is the phase operator $F$. The space of solutions $\H$
should be regarded as a \textit{real} Hilbert space, on which we are
free to impose a complex scalar product. The interpretation of the
negative-frequency solutions as antiparticles may be implemented by
taking the scalar product to be
$$
\<\psi,\phi>_F := \<\psi_+, \phi_+> + \<\phi_-, \psi_->.
$$
Any one-particle operator $A$ on~$\H$ can be written in block form:
$$
A = \twobytwo{A_{++}}{A_{+-}}{A_{-+}}{A_{--}}.
$$
Its odd and even parts can be distinguished with the phase operator:
\begin{eqnarray*}
A_+ & := & \twobytwoeven{A_{++}}{A_{--}} = \thalf(A + FAF),
\\
A_- & := & \twobytwoodd{A_{+-}}{A_{-+}} = \thalf(A - FAF)
 = \thalf F[F,A].
\end{eqnarray*}

Choose any orthonormal bases $\{\phi_k\}$ for $\H^+$ and $\{\psi_k\}$
for $\H^-$, and let $b_k := b(\phi_k)$, $d_k := d(\psi_k)$ denote
corresponding annihilation operators, together with the creation
operators $b_k^\7$, $d_k^\7$. The quantum counterpart of $A$ on Fock
space is
\begin{equation}
\mathbf{A} := b^\7 A_{++} b + b^\7 A_{+-} d^\7 + d A_{-+} b
              + \mathopen: d A_{--} d^\7 \mathclose:,
\label{eq:second-qntn}
\end{equation}
where $b^\7 A_{++} b := \sum_{j,k} b_k^\7 \<\phi_k,A_{++}\phi_j> b_j$,
$b^\7 A_{+-} d^\7 := \sum_{j,k} b_k^\7 \<\phi_k,A_{+-}\psi_j> d_j^\7$,
and so on; the last term is normal-ordered. This
``second-quantization'' rule corresponds to the infinitesimal spin
representation on Fock space, developed originally by Shale and
Stinespring.\cite{ShaleS} It is independent of the orthonormal bases
used, but makes sense only when $A_{+-}$ and $A_{+-}$ are
Hilbert--Schmidt or, equivalently, only when $[F,A]$ is
Hilbert--Schmidt. The details of this quantization recipe can be
found, for instance, in Refs.~\citelow{Polaris,Rhea}.

In $1 + 1$ dimensions, this is enough, since we can usually guarantee
that $[F,A]$ lies in the Hilbert--Schmidt class~$\L^2$ for pertinent
operators~$A$; and then $A$ is implementable on Fock space
by~(\ref{eq:second-qntn}). In other words, normal ordering is
sufficient to regularize the theory.

For higher dimensions, we seek to implement gauge transformations on
Fock space. Instead of seeking out the most general gauge potentials,
we only examine the trivial vector bundles of rank one (that is,
$\E = \A$), with gauge potentials in $\Om_D^1$. If
$A = \sum_j a_j\,[D,b_j]$, then (since $FD = DF$), we find that
$$
[F,A] = \tsum_j [F,a_j]\,[D,b_j] + a_j\,[D,[F,b_j]],
$$
and so the issue is to determine when $[F,a] \in \L^2$ for all
$a \in \A$.

The triple $(\A,\H,F)$ is called a \textit{Fredholm module} over the
algebra~$\A$. In general, such an object consists of an algebra $\A$
represented on a Hilbert space $\H$, together with a selfadjoint
operator $F$ such that $F^2 = 1$ and $[F,a]$ is compact for each
$a \in \A$. When $F$ is the phase of the operator $D$ defining a
spectral triple, these properties are clear, except possibly the
compactness of $[F,a]$. However, a straightforward spectral-theory
calculation\cite{SchroheWW} shows that, when $a^* = -a$ so that
$[F,a]$ is selfadjoint, the operator inequality
\begin{equation}
-\|[D,a]\| \,|D|^{-1} \leq [F,a] \leq \|[D,a]\| \,|D|^{-1}
\label{eq:Fcomm-vs-Dinv}
\end{equation}
holds, so that $[F,a]$ is compact since $|D|^{-1}$ is compact and
$[D,a]$ is bounded.

In $1 + 1$ quantum field theory, the Schwinger term can be written as
$$
\a(A,B) = \frac{1}{8}\, \Tr(F[F,A][F,B]),
$$
if $A$ and $B$ represent infinitesimal gauge transformations, for
which $[F,A]$ and $[F,B]$ lie in~$\L^2$. As well as being a
$2$-cocycle for Lie algebra cohomology, it is also a cyclic
$1$-cocycle.\cite{Araki} Indeed, its coboundary is
$$
b\,\a(a,b,c)
 = \frac{1}{8} \Tr(F[F,ab][F,c] - F[F,a][F,bc] + F[F,ca][F,b]) = 0.
$$

To go beyond the $1 + 1$ case, we start by noting that if
$a \in \Coo(M)$ for $M$ a spin manifold of dimension~$n$, and if $F$
is the phase of the Dirac operator, then $[F,a] \in \L^p$ for $p > n$;
this can be proved with pseudodifferential calculus. Here $\L^p$ means
the Schatten $p$-class; a compact operator $A$ lies in $\L^p$ if
$\sum_k s_k(A)^p$ converges or, equivalently, if $|A|^p$ is
traceclass. More generally, if $[F,a_j] \in \L^p$ for
$j = 1,\dots,k$, then\cite{SimonTrace}
$$
[F,a_1] [F,a_2] \dots [F,a_k] \in \L^{p/k}
$$
so that the following \textit{Chern character} makes sense, provided
that each $[F,a_j]$ lies in $\L^{n+1}$:
\begin{equation}
\tau_F(a_0,\row a1n)
 := \thalf \Tr(\chi F\,[F,a_0]\,[F,a_1]\dots [F,a_n]).
\label{eq:Chern-char}
\end{equation}
This is an $(n+1)$-linear form on~$\A$, which is cyclic ---that is,
moving $[F,a_n]$ to the position before $[F,a_0]$ changes it only by
the sign $(-1)^n$ of the corresponding cyclic permutation--- and one
can check that its coboundary vanishes: $\tau_F$ is a \textit{cyclic
\mbox{$n$-cocycle}}. We say that the Fredholm module $(\A,\H,F)$ has
``quantum dimension'' $n$ if every $[F,a]$ lies in $\L^{n+1}$.

\subsection{The quantum dimension of a noncommutative torus}

For the NC torus $\T_\th^n$, the question of UV divergences comes
down to this: is it possible to improve the implementability of gauge
transformations by showing that $[F,a] \in \L^p$ for some $p \leq n$
and all~$a \in \A$? To answer this question, we compute the quantum
dimension of a $3$-torus.

We examine the effect of $[F,a]$ on the two-dimensional subspace
spanned by $\{\psi_r^+,\psi_r^-\}$ for some fixed $r \in \Z^3$. From
(\ref{eq:DandF-block}), we know that
$F\psi_r^\pm = |r|^{-1}(\sg\.r)\psi_r^\pm$. On the other hand, if
$a = \sum_s a_s u^s$, the left multiplication operator $a$ does not
act block diagonally; instead, using (\ref{eq:NCT-prod}), we find that
$a\psi_r^\pm = \sum_s \la(s,r) a_s \psi_{r+s}^\pm$. Thus,
$$
[F,a]\psi_r^\pm = \sum_s \la(s,r) a_s\, \biggl(
 \frac{\sg\.(r+s)}{|r+s|} - \frac{\sg\.r}{|r|}\biggr)\,\psi_{r+s}^\pm.
$$
In the special case $a = u^s$, the operator $[F,u^s]^* [F,u^s]$ is
diagonal, so
\begin{eqnarray*}
[F,u^s]^* [F,u^s] \psi_r^\pm
&=& \la(r+s,s) \, \la(s,r) \biggl(
  \frac{\sg\.(r+s)}{|r+s|} - \frac{\sg\.r}{|r|}\biggr)^2 \,\psi_r^\pm
\\
&=& 2 \biggl( 1 - \frac{(r+s)\.r}{|r+s|\,|r|} \biggr) \,\psi_r^\pm.
\end{eqnarray*}
Therefore, if $p \geq 1$,
\begin{eqnarray*}
\|[F,u^s]\|_p^p
&=& \|[F,u^s]^* [F,u^s]\|_p^{p/2} = 2\.2^{p/2} \sum_{r\neq 0}
  \biggl( 1 - \frac{(r+s)\.r}{|r+s|\,|r|} \biggr)^{\!p/2}
\\
&=& 2 \sum_{r\neq 0} \biggl\{2 - 2\Bigl(1 + \frac{r\.s}{|r|^2} \Bigr)
     \Bigl(1 + 2\frac{r\.s}{|r|^2} + \frac{|s|^2}{|r|^2} \Bigr)^{-1/2}
	  \biggr\}^{p/2}
\\
&=& 2 \sum_{r\neq 0} \biggl\{ \frac{|r\x s|^2}{|r|^4}
       + O(|r|^{-3}) \biggr\}^{p/2}.
\end{eqnarray*}
We conclude that $[F,u^s] \in \L^p$ iff $\sum_{r\neq 0} |r|^{-p}$
converges, iff $\int_1^\infty \rho^{2-p} \,d\rho$ converges, if and
only if $p > 3$.

On the other hand, a similar computation\cite{Atlas} shows
that $\|[F,a]\|_4^4 < \infty$ for all $a \in \T_\th^3$; thus
$[F,a] \in \L^4$ in all cases. We conclude that the quantum dimension
of $\T_\th^3$ is greater than~$2$ but not greater than~$3$, so it
equals~$3$. Recalling now the calculation (\ref{eq:cldim-NCTthree}),
that shows that the classical dimension is also~$3$, we see that both
dimensions coincide.

It is noteworthy that the parameter $\th$ enters the above calculation
only through the cocycle $\la(s,r)$, which gets estimated by its
absolute value and ceases to matter. Therefore, the overall UV
behaviour is the same for noncommutative as for commutative tori, and
no improvement due to noncommutativity can be obtained.

\subsection{The NC index theorem prevents UV improvement}

At first, one might think that the previous ``no-go theorem'' could be
a particular feature of canonical quantization, or an accidental
property of the tori. But we may recall that Filk\cite{Filk} found the
same result by a different argument for noncommutative~$\R^4$, and
that Krajewski and Wulkenhaar\cite{KrajewskiW,KrajewskiThesis} also
found the same UV divergences for a Yang--Mills field over the
$4$-torus, by path-integral methods. There have been a few claims of
UV improvement in some models, but these either lie outside our
framework of spin geometry, or treat zero-dimensional approximations
only.

To see why it must be so, we call upon one of the deepest results in
noncommutative geometry, which establishes the relation between the
noncommutative integral defined by a generalized Dirac operator $D$
and the Chern character of its phase operator $F$. This is the
Hauptsatz of Connes, stating that both of these integrals take the
same values on ``volume forms'': see p.~308 of Ref.~\citelow{Book}.
The precise statement is that, for a geometry of classical
dimension~$n$, the following equality holds:
\begin{equation}
\thalf \Tr\bigl(\chi F\tsum_j [F,a_0^j]\,[F,a_1^j]\dots[F,a_n^j]\bigr)
 = \ncint \chi \tsum_j a_0^j\,[D,a_1^j]\dots [D,a_n^j] \,|D|^{-n},
\label{eq:Haupt-satz}
\end{equation}
whenever $\sum_j a_0^j \ox a_1^j \oxyox a_n^j$ is a Hochschild
$n$-cycle over~$\cal A$.

The proof of this theorem is a long story; this is not surprising,
because it includes the Atiyah--Singer index theorem for the case
$\A = \Coo(M)$, as is made plain in Ref.~\citelow{ConnesMNovi}.
It is also a particular case of the local index theorem in NCG
developed by Connes and Moscovici.\cite{ConnesMIndex} By
interpolating a one-parameter family of cyclic cocycles between both
sides of (\ref{eq:Haupt-satz}), one can construct a McKean--Singer
type of proof: see Chapter~10 of our Ref.~\citelow{Polaris}.

The right hand side reduces to an ordinary integral in the commutative
case. In fact, for the Dirac operator (and untwisted spin structure)
on~$\T^3$, it can be shown that
$$
\ncint a_0\,[D,a_1][D,a_2][D,a_2]\,|D|^{-3}
 = \frac{i}{3\pi^2} \int_{\T^3} a_0\,da_1 \w da_2 \w da_3.
$$
This is achieved by Connes' trace theorem,\cite{ConnesAction} which
establishes that the noncommutative integral of a pseudodifferential
operator is proportional to its Wodzicki residue, which in turn is
given by the ordinary integral of a local density. Direct computation
of the Chern character is more difficult, since $F$ is given by a
singular integral operator, and the trace is, in principle, a highly
nonlocal integral of a suitable kernel. For $\T^3$ (or $\R^3$), 
Langmann\cite{LangmannNCI} obtained
$$
\thalf \Tr\bigl(\chi F\,[F,a_0][F,a_1][F,a_2][F,a_3]\bigr)
 = \frac{i}{3\pi^2} \int_{\T^3} a_0\,da_1 \w da_2 \w da_3,
$$
which nicely corroborates the index theorem, and indeed suffices for
toral geometries.

Now, if $\cc$ is the Hochschild $n$-cycle providing the orientation of
a given spin geometry, for which $\pi_D(\cc) = \chi$, we can plug it
into the formula (\ref{eq:Haupt-satz}). Using an obvious notation, we
get
$$
\tau_F(\cc) = \ncint \chi \pi_D(\cc) \,|D|^{-n} = \ncint |D|^{-n}
 = C > 0,
$$
since $\chi^2 = 1$. If the quantum dimension were less than $n$, the
Chern character $\tau_F$ would have to be of the form
$(-2/n)\,S\tau'_F$, where $\tau'_F$ is the analogous cyclic
$(n-2)$-cocycle, and $S$ is the periodicity operator that promotes
cyclic $(n-2)$-cocycles to cyclic $n$-cocycles; this follows from the
periodicity theorem in cyclic cohomology.\cite{Book} However, another
consequence of the periodicity theorem is that promoted cyclic
cocycles are trivial in Hochschild cohomology, which means that
$\tau_F(\cc)$ would vanish. So the Hauptsatz guarantees that the
quantum dimension is at least~$n$.

On the other hand, the estimate (\ref{eq:Fcomm-vs-Dinv}) shows that
$[F,a] \in \L^p$ whenever $|D|^{-1} \in \L^p$, which happens for all
$p > n$. Thus, the quantum dimension is exactly~$n$, so it coincides
with the classical dimension in all cases. In other words, the
no-go theorem is an inescapable feature of noncommutative geometry.

\newpage

\section{Hopf algebras in noncommutative geometry}

The local index theorem of Connes and Moscovici\cite{ConnesMIndex}
expresses the character $\tau_F$ of a Fredholm module obtained from a
spectral triple $(\A,\H,D)$ as a complicated sum of noncommutative
integrals (which are ``local'') involving products of ope\-ra\-tors of
the form $[D^2,\dots[D^2[D,a]]\dots]$ and a compensating $|D|^{-m}$.
These can be computed, in principle, as residues of certain zeta
functions, but the computations turned out to be very extensive, even
for geometries of dimension~$1$. In seeking to streamline the
calculations, Connes and Moscovici identified a Hopf algebra that
governs the terms appearing in the index formula.\cite{ConnesMHopf}
This Hopf algebra, which we shall briefly describe, is commutative but
highly noncocommutative: it is not a familiar one from the literature
on quantum groups.

A short while previously, Kreimer\cite{Kreimer} had found a similar
Hopf algebra in a seemingly different context, namely, the
combinatorial structure of the Zimmermann forest formula for
renormalizing integrals corresponding to Feynman graphs! It turns out
that both Hopf algebras are closely related.\cite{ConnesKrHopf}
Subsequent work has shown that these Hopf algebra structures point to
a new and deep relationship between NCG and QFT.\cite{ConnesKrRH} We
cannot do justice to it here, so we shall merely sketch the origins of
these Kreimer--Connes--Moscovici algebras and show how they are
related.

\subsection{An example of diffeomorphism-invariant geometry}

To deal with gravity in the NCG framework, one seeks to understand the
geometrical features that are invariant under
diffeomorphisms.\cite{ConnesGravBreak} For instance, Landi and
Rovelli\cite{LandiR} have explored how the eigenvalues of the Dirac
operator provide the natural variables for such a theory.

As a first step, we can consider how to study the invariants of an
oriented manifold $M$ under the action of a subgroup $\Ga$ of the
diffeomorphism group $\Diff(M)$. The orbit space $M/\Ga$ is the leaf
space of a foliation, but could have an unpleasant topology (think of
the Kronecker foliation of a $2$-torus under rotations at an
irrational angle). In NCG, where we use the algebra instead of the
space, this problem is remedied by taking the ``crossed product''
$\A_0 := \Coo(M) \semi \Ga$ as the natural algebra of coordinates. The
crossed product is defined as the algebra generated by the functions
in $\Coo(M)$ and a set of unitaries $\set{V_\psi : \psi \in \Ga}$,
subject to the relations
$$
V_\psi^\0 h V_\psi^\7 = h^\psi,  \sepword{where}
 h^\psi(x) := h(\psi^{-1}(x)).
$$

Such an algebra is highly noncommutative, and may lack easily
constructed representations; there will often be no $\Ga$-invariant
measure that would help to build a Hilbert space ($\Ga$ could be,
say, the full diffeomorphism group). The way out of this dilemma
is to replace $M$ by the (oriented) frame bundle $F \to M$, and to
find a $\Ga$-invariant measure on~$F$. To see how this works, consider
the one-dimensional case. As usual, we take $M$ to be compact, so it
is just the circle $M = \Sf^1$ with local coordinate~$\th$. The frame
bundle is a cylinder, with vertical coordinate $y = e^{-s}$, and any
$\phi \in \Ga \subseteq \Diff^+(\Sf^1)$ acts on~$F$ by
$$
\tilde\phi(\th,s) = (\phi(\th), s - \log \phi'(\th)).
$$
Then it is easy to see that each $\tilde\phi$ preserves the measure
$e^s\,ds\,d\th$ on~$F$. (This means that $F$ can be depicted as the
funnel obtained by revolution of the graph of the logarithm.) We can
now use the Hilbert space $\H := L^2(F, e^s\,ds\,d\th)$.

The corresponding algebra $\A := \Coo_0(F) \semi \Onda\Ga$ acts
on~$\H$ and is generated by elements
$\set{fU_\psi^\7 : f \in \Coo_0(F), \psi \in \Ga}$, with the product
rule
$$
(fU_\psi^\7)(gU_\phi^\7) := f(g \circ \tilde\psi) U_{\phi\psi}^\7.
$$
The horizontal and vertical vector fields
$$
X := e^{-s} \dd{\th},  \qquad  Y := - \dd{s}
$$
serve to define an operator $D$ by solving the equation
$D|D| = Y^2 + X$; this is the $D$ that enters into the local index
computation. To compute anything explicitly, we need to know how $X$
and $Y$ interact with the unitary generators $U_\psi^\7$. It turns out
that $[Y,U_\psi^\7] = 0$, but the horizontal vector fields obey a more
complicated formula:
$$
X(fU_\psi^\7gU_\phi^\7) = (Xf)\,U_\psi^\7 gU_\phi^\7
  + fU_\psi^\7 X(gU_\phi^\7)
  + e^{-s} \frac{\psi''(\th)}{\psi'(\th)}\, fU_\psi^\7 Y(gU_\phi^\7).
$$
Write $\la_1(fU_\psi^\7)
 := \bigl(e^{-s} \psi''(\th)/\psi'(\th) \bigr) fU_\psi^\7$. This is a
derivation of the algebra~$\A$. The previous formula can now be
abbreviated as
\begin{equation}
X(ab) = X(a)\,b + a\,X(b) + \la_1(a)\,Y(b),  \qquad  a,b \in \A.
\label{eq:Xder-not}
\end{equation}

It is obvious that $[Y,X] = X$ as vector fields on~$F$, and this
relation transfers to their action on~$\A$; moreover,
$[Y,\la_1] = \la_1$. But $X$ is not quite a derivation and $X$, $Y$
and~$\la_1$ do not close to a Lie algebra acting on~$\A$. In fact, if
$\la_2 := [X,\la_1]$, we find that
$$
\la_2(fU_\psi^\7)
 = e^{-2s} \frac{\psi'\psi'''- {\psi''}^2}{{\psi'}^2}\, fU_\psi^\7.
$$
In general, if we introduce
$$
\la_n(fU_\psi^\7)
 := e^{-ns} \pd{^n}{\th^n} \bigl(\log \psi'(\th)\bigr)\, fU_\psi^\7,
$$
then $X,Y,\la_1,\dots,\la_n,\dots$ closes to a Lie algebra, with the
commutation relations
$$
[Y,X] = X,  \quad  [X,\la_n] = \la_{n+1},  \quad
[Y,\la_n] = n\la_n,  \quad  [\la_m,\la_n] = 0.
$$

We can rewrite (\ref{eq:Xder-not}) and the Leibniz rule for~$Y$ in the
language of coproducts:
\begin{eqnarray*}
\Dl Y & = & Y \ox 1 + 1 \ox Y,  \\
\Dl X & = & X \ox 1 + 1 \ox X + \la_1 \ox Y.
\end{eqnarray*}
The analogous rules for the $\la_1$, $\la_2$, $\la_3$ may be found
from these and the commutation rules, bearing in mind that $\Dl$ is a
homomorphism of algebras:
\begin{eqnarray}
\Dl \la_1 &=& \la_1 \ox 1 + 1 \ox \la_1,
\nonumber \\
\Dl \la_2 &=& \la_2 \ox 1 + 1 \ox \la_2 + \la_1 \ox \la_1,
\nonumber \\
\Dl \la_3 &=& \la_3 \ox 1 + 1 \ox \la_3 + 3\la_1 \ox \la_2
   + \la_2 \ox \la_1 + \la_1^2 \ox \la_1.
\label{eq:coprod-CM}
\end{eqnarray}
We can continue recursively, using $\Dl\la_{n+1} = [\Dl X, \Dl\la_n]$.
{}From (\ref{eq:coprod-CM}) it is clear that the elements $\la_n$
(without $X$ or~$Y$) generate a Hopf algebra that is commutative but
by no means cocommutative.

Any Hopf algebra comes equipped with an antipode~$S$, which is the
unique linear map from the Hopf algebra into itself such that both
$m(S \ox \id)\Dl$ and $m(\id \ox S)\Dl$ are equal to the map taking
$1$ to~$1$ and all other generators to~$0$. (Here $m$ denotes the
algebra product.) More explicitly, if
$\Dl \a := \sum_j \a'_j \ox \a''_j$ is the coproduct of a
nontrivial generator $\a$, then
$\sum_j S(\a'_j) \a''_j = \sum_j \a'_j S(\a''_j) = 0$. With the
coproduct and product in hand, the antipode can be determined; see
Refs.~\citelow{ConnesMHopf,Polaris} or Ref.~\citelow{Ananke} for 
the details. For example,
$$
S(\la_1) = - \la_1, \quad  S(\la_2) = - \la_2 + \la_1^2, \quad
S(\la_3) = - \la_3 + 4\la_1\la_2 - 2\la_1^3.
$$

With these tools, one can continue to compute the index formula in
low-dimensional cases.\cite{ConnesMHopf}

\subsection{Nested subdivergences and the forest formula}

And now for something completely different. Suppose that we wish to
deal with a multiloop Feynman graph with superficially divergent
subgraphs, which we hope to renormalize by subtracting appropriate
counterterms. If there are no overlapping divergences, the
subdivergences form a family of subgraphs that are either nested or
disjoint; such a family is called a forest. The counterterms
may be assigned by Zimmermann's forest formula,\cite{Zimmermann} which
is a sum over all forests in the given diagram and constitutes a
recursive rule for the several levels of nesting.
Kreimer\cite{Kreimer} discovered that Zimmermann's procedure is
encoded in a Hopf algebra, whose elements are called ``rooted
trees''.\cite{ConnesKrHopf}

Given a diagram with only nested or disjoint subdivergences, the
\textit{root} of the tree, depicted at the top, represents the full
diagram. The leaves of the tree are subdivergences that include no
proper subdivergences, and ascending links indicate the intermediate
nestings. Nodes that are linked only through a higher node represent
disjoint subdivergences. Here are all the rooted trees with four
nodes:
$$
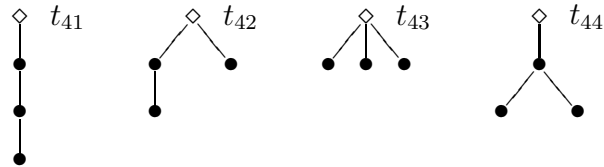
\xy 0;<18pt,0pt>: *{t_{41}} ;
(0,0)*{\5} ; (0,-1)*{\8} **@{-} ;
(0,-2)*{\8} **@{-} ; (0,-3)*{\8} **@{-}
\endxy
\qquad
\xy 0;<18pt,0pt>: *{t_{42}} ;
(0,0)*{\5} ; (.8,-1)*{\8} **@{-} , (-.8,-1)*{\8} **@{-} ;
(-.8,-2)*{\8}
**@{-}
\endxy
\qquad
\xy 0;<18pt,0pt>: *{t_{43}} ; (0,0)*{\5} ;
(-.8,-1)*{\8} **@{-} , (0,-1)*{\8} **@{-} , (.8,-1)*{\8} **@{-}
\endxy
\qquad
\xy 0;<18pt,0pt>: *{t_{44}} ;
(0,0)*{\5} ; (0,-1)*{\8} **@{-} ;
(-.8,-2)*{\8} **@{-} , (.8,-2)*{\8} **@{-}
\endxy
$$

These rooted trees generate a commutative \textit{algebra}, whose unit
$1$ corresponds to the empty tree. The product is denotes by
juxtaposition, and the sum is a formal one (it corresponds to a sum of
integrals for several Feynman graphs). We make it a \textit{Hopf}
algebra by introducing a coproduct and identifying an antipode.

To get the coproduct, we need to see how the tree may be cut by
lopping off one or more branches from the root part (which we call the
\textit{trunk}), without ever cutting a piece already separated from
the root. For the rooted tree denoted $t_{42}$, here are the allowable
cuts:
$$
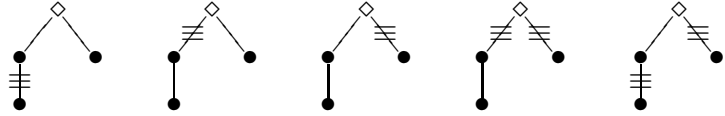
\xy 0;<18pt,0pt>:
(0,0)*{\5} ; (.8,-1)*{\8} **@{-} ,
(-.8,-1)*{\8}  **@{-} ; (-.8,-2)*{\8} **@{-} ?*{\equiv}
\endxy
\qquad
\xy 0;<18pt,0pt>:
(0,0)*{\5} ; (.8,-1)*{\8} **@{-} ,
(-.8,-1)*{\8}  **@{-} ?*{\equiv} ?(1) ; (-.8,-2)*{\8}
**@{-}
\endxy
\qquad
\xy 0;<18pt,0pt>:
(0,0)*{\5} ; (.8,-1)*{\8} **@{-} ?*{\equiv} ?(1) ,
(-.8,-1)*{\8}  **@{-} ; (-.8,-2)*{\8} **@{-}
\endxy
\qquad
\xy 0;<18pt,0pt>:
(0,0)*{\5} ; (.8,-1)*{\8} **@{-} ?*{\equiv} ?(1) ,
(-.8,-1)*{\8}  **@{-} ?*{\equiv} ?(1) ; (-.8,-2)*{\8} **@{-}
\endxy
\qquad
\xy 0;<18pt,0pt>:
(0,0)*{\5} ; (.8,-1)*{\8} **@{-} ?*{\equiv} ?(1) ,
(-.8,-1)*{\8}  **@{-} ; (-.8,-2)*{\8} **@{-} ?*{\equiv}
\endxy
$$

For each allowable cut $c$ of a rooted tree~$T$, we denote the trunk
by $R_c(T)$, and the product of the pruned branches by $P_c(T)$. The
coproduct satisfies $\Dl(1) := 1 \ox 1$, and is given on the
nontrivial generators by
$$
\Dl T := T \ox 1 + 1 \ox T + \sum_c P_c(T) \ox R_c(T).
$$
The first two terms may be absorbed in the sum by adding an ``empty
cut'' whose trunk is the whole tree, and a ``full cut'' that prunes
the whole tree. Here, for example, is $\Dl(t_{42})$:
$$
\Dl\biggl( \vcenter{
 \xy 0;<10pt,0pt>: (0,0)*{\5} ; (.8,-1)*{\8} **@{-} ,
  (-.8,-1)*{\8}  **@{-} ; (-.8,-2)*{\8} **@{-} \endxy} \biggr)
= \vcenter{ \xy 0;<10pt,0pt>: (0,0)*{\5} ; (.8,-1)*{\8} **@{-} ,
     (-.8,-1)*{\8}  **@{-} ; (-.8,-2)*{\8} **@{-} \endxy} \ox 1
 + 1 \ox \vcenter{\xy 0;<10pt,0pt>: (0,0)*{\5} ; (.8,-1)*{\8} **@{-},
            (-.8,-1)*{\8}  **@{-} ; (-.8,-2)*{\8} **@{-} \endxy}
 + \5 \ox \vcenter{ \xy 0;<10pt,0pt>: (0,0)*{\5} ;
                    (.8,-1)*{\8} **@{-} , (-.8,-1)*{\8} **@{-} \endxy}
 + \vcenter{\xy 0;<10pt,0pt>: (0,0)*{\5}; (0,-1)*{\8} **@{-} \endxy}
   \ox\vcenter{\xy 0;<10pt,0pt>: (0,0)*{\5}; (0,-1)*{\8} **@{-}\endxy}
 + \5 \ox \vcenter{\xy 0;<10pt,0pt>: (0,0)*{\5}; (0,-1)*{\8} **@{-} ;
                                     (0,-2)*{\8} **@{-} \endxy}
 + \vcenter{\xy 0;<10pt,0pt>: (0,0)*{\5}; (0,-1)*{\8} **@{-} \endxy}
     \5 \ox \5
 +  \5\5  \ox
   \vcenter{\xy 0;<10pt,0pt>: (0,0)*{\5}; (0,-1)*{\8} **@{-} \endxy}
$$

To determine the antipode, note first that
$1 = m(S \ox \id)(1 \ox 1) = S(1)$, and that
$m(S \ox \id)\Dl(T) = S(T) + T + \sum_c S(P_c(T))\, R_c(T)$, and thus
$$
S(T) = - T - \sum_c S(P_c(T))\, R_c(T).
$$
This recursive recipe for $S$ turns out to give precisely the forest
formula! That is proved in Ref.~\citelow{Ananke}. Thus the lore of
counterterms may be reduced to understanding the recipe for the above
coproduct of trees.

\subsection{How both Hopf algebras are related}

To find the relationship between Kreimer's Hopf algebra and that of
Connes and Moscovici, we must do a little gardening. There is a unique
rooted tree with one node (the solitary root, called $t_1$) and only
one with two nodes (call it $t_2$); there are two rooted trees with
three nodes, called $t_{31}$ (three nodes in a chain) and $t_{32}$
(two leaves sprouting from the root), and we have already seen the
four rooted trees with four nodes. There is an easy way to produce new
trees from old,\cite{ConnesKrHopf} called ``natural growth'': for any
tree $T$, let $N(T)$ be the \textit{sum} of all the trees formed by
adding one new branch and leaf at \textit{each} node of~$T$. Thus
$N(t_{31}) = t_{41} + t_{42} + t_{43}$,
$N(t_{32}) = 2t_{42} + t_{44}$, and so on.

Introduce $\dl_1 := t_1$, $\dl_2 := N(t_1) = t_2$,
$\dl_3 := N(t_2)  = t_{31} + t_{32}$, and in general
$\dl_{n+1} := N(\dl_n)$. For example,
$\dl_4 = t_{41} + 3t_{42} + t_{43} + t_{44}$. If we compute the
coproducts of these sums of trees, we find that
\begin{eqnarray*}
\Dl \dl_1 &=& \dl_1 \ox 1 + 1 \ox \dl_1,
\nonumber \\
\Dl \dl_2 &=& \dl_2 \ox 1 + 1 \ox \dl_2 + \dl_1 \ox \dl_1,
\nonumber \\
\Dl \dl_3 &=& \dl_3 \ox 1 + 1 \ox \dl_3 + 3\dl_1 \ox \dl_2
   + \dl_2 \ox \dl_1 + \dl_1^2 \ox \dl_1,
\end{eqnarray*}
and so on. These are exactly the same formulas
as~(\ref{eq:coprod-CM})!

Why so? The connection lies in observing that the natural growth
operator satisfies a Leibniz rule:
$$
N(T_1T_2) = N(T_1) T_2 + T_1 N(T_2),
$$
since, given two juxtaposed trees $T_1$ and $T_2$, we may hang the
extra node on either one. We may express this by forming the (abelian)
Lie algebra generated by the trees and introducing an extra generator
$X$ by declaring that $[X,T] := N(T)$. Moreover, the number of nodes
$\# T$ in a tree~$T$ gives a $\Z$-grading on the algebra of trees,
since $\#(T_1 T_2) = \# T_1 + \# T_2$. We can then add another
generator $Y$ to the Lie algebra by declaring that
$[Y,T] := (\# T)\,T$. Finally, observe that
\begin{eqnarray*}
[[Y,X],T] &=& [[Y,T],X] + [Y,[X,T]] = (\#T)\,[T,X] + [Y,N(T)]
\\
& = & -(\#T)\,N(T) + (\#T + 1)\,N(T) = N(T) = [X,T],
\end{eqnarray*}
so that the Lie algebra closes with $[Y,X] = X$.

We can compute the coproduct $\Dl(N(T))$, too. All we have to do is
to grow an extra leaf on~$T$ and then cut the resulting trees in every
allowable way. If the new branch is not cut in this process, then it
belongs to either a pruned branch or to the trunk that remains after a
cut has been made on the original tree~$T$; this amounts to
$(N \ox \id)\Dl(T) + (\id \ox N)\Dl(T)$. On the other hand, if the new
branch is cut, the new leaf contributes a solitary node $\dl_1$ to
$P_c$; the new leaf must have been attached to the trunk $R_c(T)$ at
any one of the latter's nodes. Since $(\#R_c)R_c = [Y,R_c]$, the
terms wherein the new leaf is cut amount to $[\dl_1 \ox Y, \Dl(T)]$.
In total,
$$
\Dl(N(T))
 = (N \ox \id) \Dl(T) + (\id \ox N) \Dl(T) + [\dl_1 \ox Y, \Dl T].
$$
Thus, since $\Dl[X,T] = [\Dl(X), \Dl(T)]$ must hold, we get
$$
\Dl(X) = X \ox 1 + 1 \ox X + \dl_1 \ox Y.
$$
Also, it is easy to see that $\Dl(Y) = Y \ox 1 + 1 \ox Y$.

The conclusion is that, with these extra generators $X$ and $Y$, the
Hopf subalgebra generated by $X$, $Y$ and the various $\dl_n$ is
isomorphic to the Connes--Moscovici algebra.

The moral of the story is that Hopf algebras provide a new and
useful entry point for noncommutative geometry into the business
of renormalization. The hope that this will shed new light on QFT
can already be justified.\cite{ConnesKrRH,ConnesKrQFT}

\nonumsection{Acknowledgements}
 
I am much indebted to Jos\'e M. Gracia-Bond\'{\i}a for enlightening
comments, and to H\'ector Figueroa and Franciscus Vanhecke for helpful
discussions. Support from the Vicerrector\'{\i}a de Investigaci\'on de
la Universidad de Costa Rica is acknowledged.

\nonumsection{References}

\end{document}